# Structure, antiferromagnetism and superconductivity of the layered iron arsenide NaFeAs†

Dinah R. Parker,[a] Michael J. Pitcher,[a] Peter J Baker,[b] Isabel Franke,[b] Tom Lancaster,[b] Stephen J. Blundell[b] and Simon J. Clarke*[a]

**A new layered iron arsenide NaFeAs isostructural with the superconducting lithium analogue, displays evidence for the coexistence of superconductivity and magnetic ordering.**

Following the discovery of superconductivity in fluorine-doped LaFeAsO ($T_c$ = 26 K),[1] a new class of superconductor has emerged consisting of compounds containing iron pnictide or chalcogenide layers. LaFeAsO takes the ZrSiCuAs structure (filled PbFCl type) with anti-PbO-type FeAs layers in which Fe is tetrahedrally coordinated by As. Replacement of some oxide by an equal amount of fluoride[1] or vacancies[2] (electron doping) induces superconductivity. Substitution of La by other rare earths[2–8] has resulted in $T_c$ as high as 55 K in SmFeAsO$_{1-x}$.[2] Superconductivity at up to 38 K has also been discovered in derivatives of $A$Fe$_2$As$_2$ ($A$ = Ca, Sr, Ba) with the ThCr$_2$Si$_2$ structure.[9–12] Recently, we and others identified LiFeAs, with the PbFCl structure type and also containing anti-PbO-type FeAs layers, as a superconductor below 18 K when stoichiometric.[13,14] Superconductivity has also been reported in α-FeSe consisting of anti-PbO-type layers.[15] The mechanism for superconductivity in all these compounds is unclear; like the high $T_c$ layered cuprate superconductors, the iron pnictide and chalcogenide systems seem to be non-BCS superconductors close in composition to magnetically ordered phases. Current research is geared towards investigating correlations between the occurrence of superconductivity and the value of $T_c$, and the electron count and geometric parameters (e.g. Fe−As distances and As−Fe−As angles) by making substitutions at various crystallographic sites. Here we report a new compound, NaFeAs, isostructural with LiFeAs which appears to contain a mixture of superconducting (below 9 K) and magnetically ordered (below 40 K) portions.

NaFeAs was synthesised‡ by the reaction of stoichiometric quantities of elemental reagents. Fe and As powders were ground together and added to pieces of Na in a tantalum tube which was then sealed by welding under 1 atm argon gas; the mixture was heated at 800 °C for 2 days. Preliminary characterisation of the resulting air sensitive product by powder X-ray diffraction (PXRD) revealed apparently phase pure NaFeAs, isostructural with LiFeAs.[13,14] This sample was used in its entirety for neutron powder diffraction (NPD) measurements for detailed structural analysis (POLARIS, ISIS Facility, UK)‡ and portions of this sample were used for all the other measurements reported here. NaFeAs adopts the tetragonal space group $P4/nmm$ ($a$ = 3.9494(2) Å and $c$ = 7.0396(8) Å at 298 K); refinement results (Fig. 1) and crystallographic parameters are included in the electronic supplementary information (ESI)†. Refinement of fractional site occupancies produced a stoichiometry Na$_{0.98(1)}$FeAs$_{0.98(1)}$ (with the Fe occupancy fixed) which shows that there is no significant deviation from the 1:1:1 stoichiometry compared with the uncertainty in the refined parameters. No impurity phases were detected by NPD.

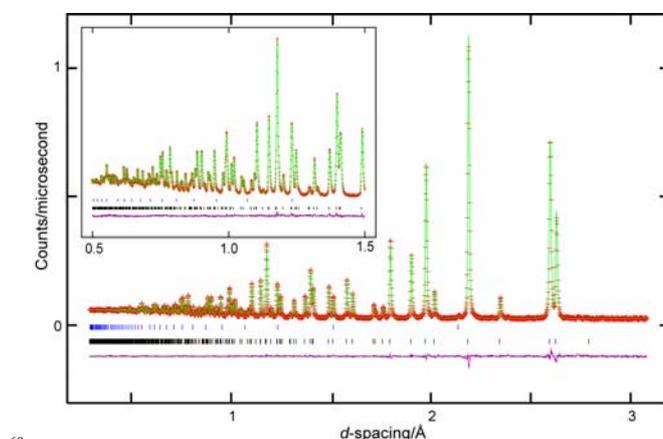

**Fig. 1** Rietveld refinement against POLARIS data (145° data bank) at 298 K. The data (red points), fit (green line) and difference (purple line) are shown. Tick marks indicate reflections for NaFeAs (black) and the vanadium container (blue); Inset: detail of the region 0.5 to 1.5 Å.

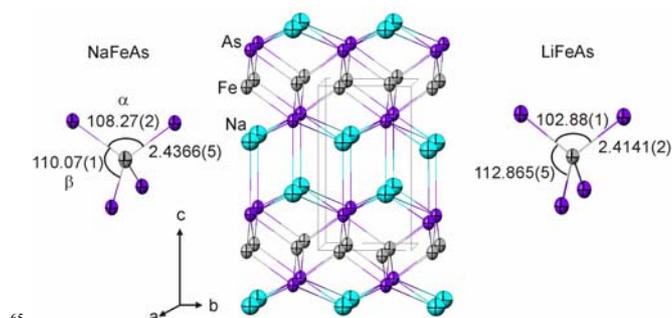

**Fig. 2** Centre: Crystal structure of NaFeAs (see text for structure description). Left: Bond distances (Å) and angles (°) in the FeAs$_4$ tetrahedron of NaFeAs at 298 K (Table 1). Right: corresponding dimensions of the FeAs$_4$ tetrahedron of isostructural LiFeAs (ref. 13). 99 % anisotropic displacement ellipsoids are shown.

The crystal structure of NaFeAs is similar to that of LiFeAs and is isotypic with those of PbFCl and Cu$_2$Sb. Fe and As are arranged in anti-PbO-type layers (see Fig. 2) with Na located between the layers in square-based pyramidal coordination by As. In the FeAs layers As ions form a distorted tetrahedral arrangement around the Fe ions, giving rise to two distinct

As−Fe−As bond angles with multiplicities of two and four which we refer to as α and β (see Fig. 2) respectively, following ref. 4. The values of α and β and the Fe−As bond length are compared in Table 1 for LiFeAs and NaFeAs. In LiFeAs, which superconducts below 18 K, α and β are 102.9° (×2) and 112.9° (×4) respectively at room temperature, giving a tetrahedron which is the most compressed in the basal plane of that of any of the iron arsenide superconductors. In NaFeAs the tetrahedron is almost regular with α and β equal to 108.2° (×2) and 110.1° (×4) respectively and the angles are closer to the values observed in the layered iron arsenides which show the highest values of $T_c$ when their electron count is optimised.[4] Although the FeAs$_4$ tetrahedra are relatively compressed along the c-axis compared with the situation in LiFeAs, the c/a ratio of NaFeAs at room temperature of 1.782 is larger than that of LiFeAs (c/a = 1.683) as a consequence of accommodating the larger Na$^+$ ion between the layers.

The NPD pattern measured on POLARIS at 2.5 K,† revealed no magnetic Bragg peaks above the background, nor was any structural transition or peak broadening evident.

**Table 1** Fe−As bond lengths and As−Fe−As angles in NaFeAs and LiFeAs,[13] compared at room temperature and below 10 K.

| | α / ° * | β / ° * | Fe−As distance / Å |
|---|---|---|---|
| LiFeAs (295 K)[14] | 102.88(1) | 112.865(5) | 2.4141(2) |
| LiFeAs (6.5 K)[14] | 103.30(1) | 112.643(6) | 2.4035(2) |
| NaFeAs (295 K) | 108.27(2) | 110.07(1) | 2.4366(5) |
| NaFeAs (2.5 K) | 108.74(1) | 109.837(6) | 2.4281(2) |

* α (multiplicity 2) and β (multiplicity 4) are defined in Fig. 2.

Fig. 3 shows the DC magnetic susceptibility versus temperature of the same sample of NaFeAs as was used for the structural analysis. The susceptibility measured in an applied field of 50 Oe shows evidence for superconducting behaviour with the onset of diamagnetism below about 9 K. However the transition to a superconducting regime is broad in temperature and the value of the zero-field-cooled susceptibility at 2 K is about 10 % of that expected for a 100 % superconducting volume fraction. The plot of magnetisation versus applied field at 300 K,† revealed the presence of 0.08 % by mass of metallic Fe (much less than can reasonably be detected by diffraction methods). The data in Fig. 3 were corrected for this impurity using the measured susceptibility of iron under the conditions used for measuring the NaFeAs sample.† This small amount of elemental iron is not sufficient to account for the small value of the diamagnetic signal. Chu and co-workers[16] have recently reported very similar behaviour (i.e. 10 % diamagnetism) in a single crystal of stoichiometric NaFeAs.

In order to probe the bulk behaviour of this compound more closely, muon-spin rotation (μSR) measurements were performed using the GPS spectrometer at the Paul Scherrer Institute, Switzerland. The powder was tightly sealed in an Ag foil packet which was sealed under argon for transport to the spectrometer. The outside of the Ag foil packet was exposed to air for less than two minutes during loading into the He atmosphere of the cryostat. Above 45 K the muon decay asymmetry has an exponentially-relaxing form typical for a paramagnet. Oscillations in the muon decay asymmetry

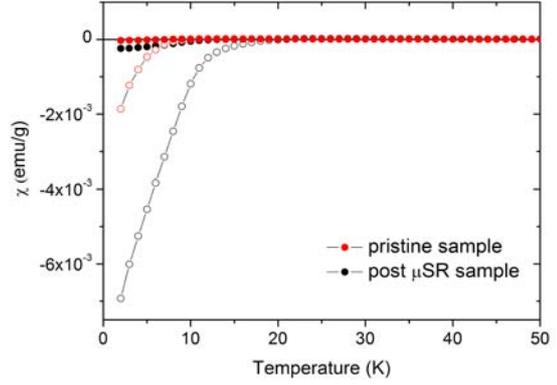

**Fig. 3** Zero-field-cooled (ZFC) (open circles) and field-cooled (FC) (closed circles) susceptibilities versus temperature for NaFeAs measured in an applied field of 50 Oe. The red symbols show the behaviour of the pristine material. The black symbols show the behaviour of a portion of the same sample which had unavoidably received brief air exposure after the muon-spin rotation experiments.

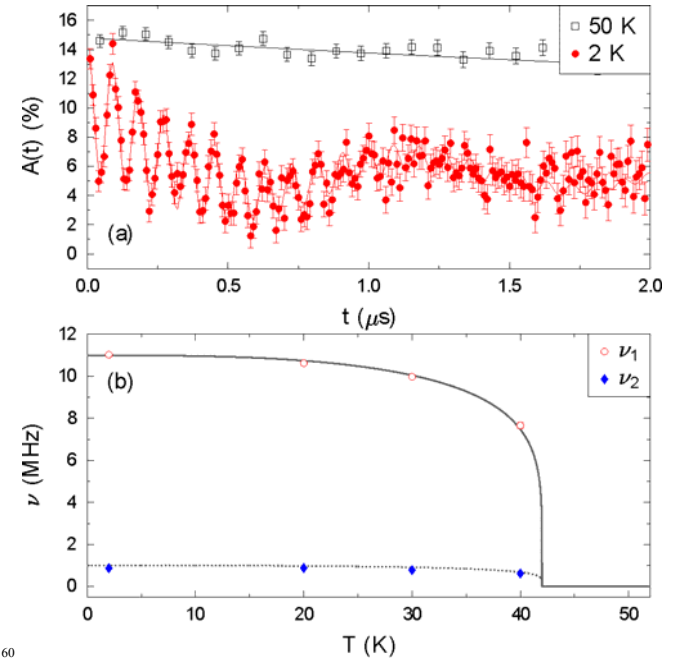

**Fig. 4.** (a) Muon decay asymmetry spectra below (2 K) and above (50 K) the magnetic ordering transition. (b) Oscillation frequencies extracted from the muon asymmetry spectra. The lines are guides to the eye based on the temperature dependence of the corresponding frequencies in SrFeAsF.[17]

(Fig. 4(a)) were observed at all temperatures below 40 K and indicate that the sample is antiferromagnetically ordered throughout at least 90% of its volume. Repeating the experiment on a second sample from the same batch gave the same behaviour. The two oscillation frequencies, corresponding to two magnetically inequivalent muon stopping sites, are plotted as functions of temperature in Fig. 4(b). The upper frequency accounts for 80% of the oscillating amplitude. Both frequencies are approximately half those observed in LaFeAsO and SrFeAsF,[17] suggesting an approximate ordered magnetic moment of the Fe atoms in

NaFeAs of 0.1−0.2 $\mu_B$, which would not be expected to lead to magnetic Bragg peaks observable above the background in our NPD measurements, nor lead to any observable anomaly in the susceptibility measurements. Transverse-field μSR measurements revealed no broadening of the spectra due to superconductivity, placing an upper limit of ~20% on the superconducting volume fraction, consistent with the magnetometry reported here and by Chu and co-workers.[16] On removal from the muon beam the Ag foil packet containing the sample experienced further air exposure, but PXRD measurements† suggested no major degradation of the sample on brief exposure to air. However, magnetometry measurements performed on the sample after the μSR measurements showed that the onset of diamagnetism now occurred at a significantly higher temperature (~ 18 K) than in the pristine material (Fig. 3) and that the diamagnetic signal at 2 K had increased to about 30 % of that expected for a bulk superconductor. This behaviour is consistent with that reported by Chu and co-workers[16] who ascribe the change in superconducting properties on air exposure to oxidative deintercalation of sodium from the compound.

Our physical property measurements show that NaFeAs which is stoichiometric within the uncertainty of the neutron diffraction measurements is not a bulk superconductor, in apparent contrast to LiFeAs.[13,14,18] The μSR experiment probes the bulk of the sample, demonstrating that magnetic order occurs throughout most of its volume below ~42 K, and places an upper limit of about 20% on the superconducting volume fraction. We see no evidence at the resolution of our neutron diffractometer for a structural phase transition associated with magnetic ordering similar to the pronounced tetragonal to orthorhombic distortions observed in LnFeAsO and $A$Fe$_2$As$_2$.[9] The magnetometry measurements reported here and by others[16] show, however, that there is a portion of the sample (about 10 %) which is superconducting at low temperatures. This result could be due to either (i) a superconducting, sodium-deficient surface layer, consistent with the results of Chu and co-workers[16] which suggest that Na-deficient bulk materials exist and have higher superconducting volume fractions than stoichiometric material, or (ii) a microscopic coexistence between magnetic and superconducting states within the stoichiometric NaFeAs system, analogous to the behaviour observed in SmFeAsO$_{1-x}$F$_x$.[19] Further experiments are in progress to compare LiFeAs and NaFeAs with the other families of layered iron pnictide and chalcogenide superconductors and to correlate the electronic and magnetic properties of NaFeAs derivatives with composition and structural parameters.

## Notes and references

[a] *Department of Chemistry, University of Oxford, Inorganic Chemistry Laboratory, South Parks Road, Oxford, OX1 3QR, UK. Fax: 44 1865 272690; Tel: 44 1865 272600; E-mail: simon.clarke@chem.ox.ac.uk*
[b] *Department of Physics, University of Oxford, Clarendon Laboratory, Parks Road, Oxford, OX1 3PU, UK.*

† Electronic Supplementary Information (ESI) available: [Tables of refinement results, refined crystallographic parameters and bond lengths; Rietveld refinement against 2.5 K NPD pattern of NaFeAs; plots showing the correction of the susceptibility for a small iron impurity; pre- and post-μSR PXRD data]. See DOI: 10.1039/b000000x/

‡ *Materials*: Na lump (BDH 99.9%), Fe powder (ALFA, 99.998 %), As pieces (ALFA, 99.9999 %). Laboratory PXRD data were collected using a PANalytical X'Pert PRO diffractometer operating with Cu$K\alpha_1$ radiation in Bragg-Brentano geometry with samples sealed inside air-tight cells. Time-of-flight PND data were collected on 1.5 g of sample sealed in a 6 mm diameter vanadium cylinder using the POLARIS diffractometer, ISIS, UK: detector banks at 35°, 90° and 145° 2$\theta$ (0.5 < d < 8.15 Å).

*Crystal data*: NaFeAs: NPD Rietveld refinement (GSAS[20]): $T$ = 295 K, tetragonal, space group $P4/nmm$ (no. 129), $a$ = 3.9494(2) Å, $c$ = 7.0396(8) Å, $V$ = 109.804(1) Å$^3$, $c/a$ = 1.782, $Z$ = 2; Na (0.25 0.25 0.64602(7)), Fe (0.75 0.25 0), As (0.25 0.25 0.20278(3)); $wR_p$ (35°, 90° and 145° banks combined) = 0.0209, $\chi^2$ = 1.523.

NaFeAs: NPD refinement: $T$ = 2.5 K, tetragonal, space group $P4/nmm$ (no. 129), $a$ = 3.94729(2) Å, $c$ = 6.99112(6) Å, $V$ = 108.930(1) Å$^3$, $c/a$ = 1.771, $Z$ = 2; Na (0.25 0.25 0.64673(8)), Fe (0.75 0.25 0), As (0.25 0.25 0.20234(4)); $wR_p$ (35°, 90° and 145° banks) = 0.0129, $\chi^2$ = 1.598.

# Structure, antiferromagnetism and superconductivity of the layered iron arsenide NaFeAs

Dinah R. Parker, Michael J. Pitcher, Peter J Baker, Isabel Franke, Tom Lancaster, Stephen J. Blundell and Simon J. Clarke*

**Table S1.** Summary of refinement of NaFeAs against Powder Neutron Diffraction (PND) data at 295 K, measured on the POLARIS instrument, ISIS facility, UK.

| *Space Group* | P4/nmm |
|---|---|
| $a$ / Å | 3.9494(2) |
| $c$ / Å | 7.0396(8) |
| $V$ / Å$^3$ | 109.804(1) |
| $R_{wp}$ | 0.0209 |
| $\chi^2$ | 1.523 |

**Table S2.** Refined atomic parameters for NaFeAs at 295 K, from data measured on the POLARIS instrument, ISIS facility, UK.

| Atom | Site | $x$ | $y$ | $z$ | $U_{equiv}$ / Å$^2$ × 100 |
|---|---|---|---|---|---|
| Fe | 2$a$ | 0.75 | 0.25 | 0 | 0.66(1) |
| Na | 2$c$ | 0.25 | 0.25 | 0.64602(7) | 1.31(1) |
| As | 2$c$ | 0.25 | 0.25 | 0.20278(3) | 0.67(1) |

**Table S4.** Refined anisotropic displacement parameters for NaFeAs from data measured at 295 K on the POLARIS instrument, ISIS facility, UK.

| Atom | $U_{11} = U_{22}$ /Å$^2$ × 100 | $U_{33}$ //Å$^2$ × 100 |
|---|---|---|
| Fe | 0.546(5) | 0.899(8) |
| Na | 1.315(9) | 1.317(9) |
| As | 0.577(7) | 0.856(8) |

**Table S5.** Summary of refinement of NaFeAs against Powder Neutron Diffraction (PND) data at 2.5 K, measured on the POLARIS instrument, ISIS facility, UK.

| *Space Group* | P4/nmm |
|---|---|
| $a$ / Å | 3.94729(2) |
| $c$ / Å | 6.99112(6) |
| $V$ / Å$^3$ | 108.930(1) |
| $R_{wp}$ | 0.0129 |
| $\chi^2$ | 1.598 |

**Table S6.** Refined atomic parameters for NaFeAs at 2.5 K, from data measured on the POLARIS instrument, ISIS facility, UK.

| Atom | Site | $x$ | $y$ | $z$ | $U_{equiv}$ / Å$^2$ × 100 |
|------|------|------|------|------------|----------------|
| Fe | 2a | 0.75 | 0.25 | 0 | 0.19(1) |
| Na | 2c | 0.25 | 0.25 | 0.64673(8) | 0.54(3) |
| As | 2c | 0.25 | 0.25 | 0.20234(4) | 0.22(2) |

**Table S7.** Refined anisotropic displacement parameters for NaFeAs from data measured at 2.5 K on the POLARIS instrument, ISIS facility, UK.

| Atom | $U_{11} = U_{22}$ /Å$^2$ × 100 | $U_{33}$ //Å$^2$ × 100 |
|------|------|------|
| Fe | 0.134(4) | 0.317(8) |
| Na | 0.429(9) | 0.753(9) |
| As | 0.184(7) | 0.298(9) |

**Table S8.** Refined bond lengths (Å) and angles (degrees) for NaFeAs at 295 K and 2.5 K from data measured on the POLARIS instrument, ISIS facility, UK.

|  | 295 K | 2.5 K |
|------|------|------|
| Fe−As [4][a] | 2.4366(5) | 2.4281(2) |
| Na−As [4] | 2.9886(7) | 2.9842(2) |
| Na−As [1] | 3.1203(9) | 3.1063(6) |
| Fe−Fe [4] | 2.7927(7) | 2.79115(3) |
| Na−Fe [4] | 3.1795(7) | 3.1620(5) |
| Na-Na [4] | 3.4678(9) | 3.4633(7) |
|  |  |  |
| As−Fe−As [2] | 108.27(2) | 108.74(1) |
| As−Fe−As [4] | 110.07(1) | 109.837(6) |
| As−Na−As [4] | 82.713(8) | 82.808(8) |
| As−Na−As[4] | 110.86(1) | 110.72(1) |

[a] The number in square brackets indicates the number of symmetry equivalent bond lengths and angles

**Figure S1.** Rietveld refinements for NaFeAs at 2.5 K, measured on the POLARIS instrument, ISIS, UK. 145º bank. (note: peaks at ~ 2.1 Å from vanadium sample holder and cryostat)

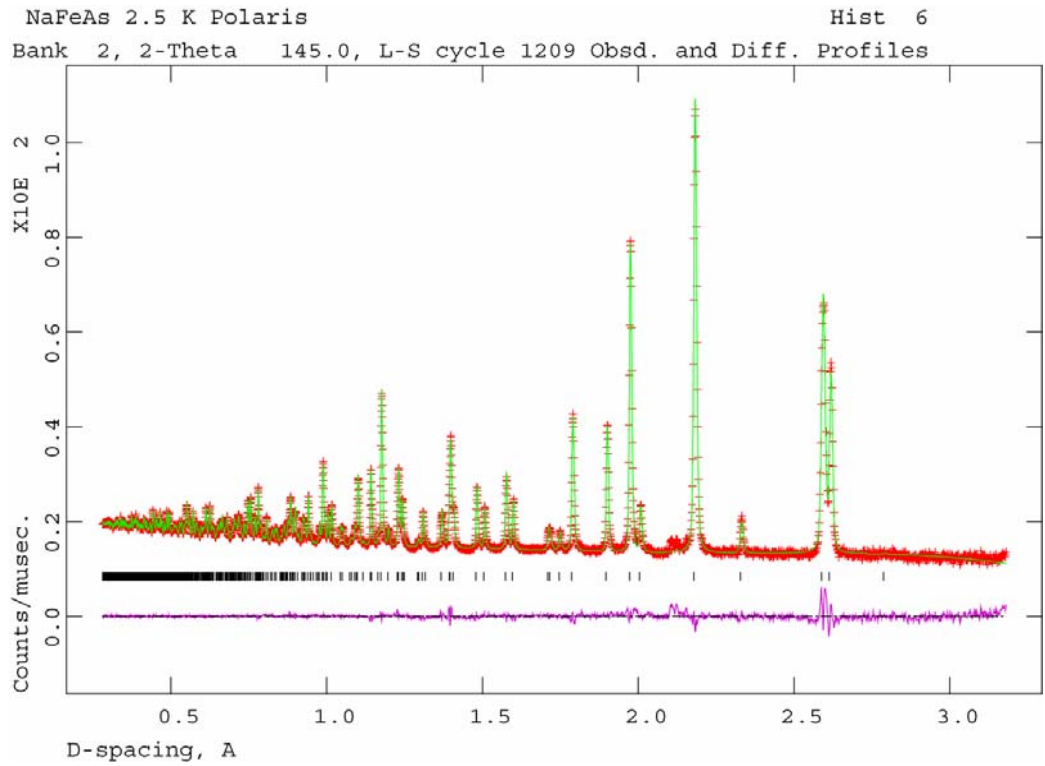

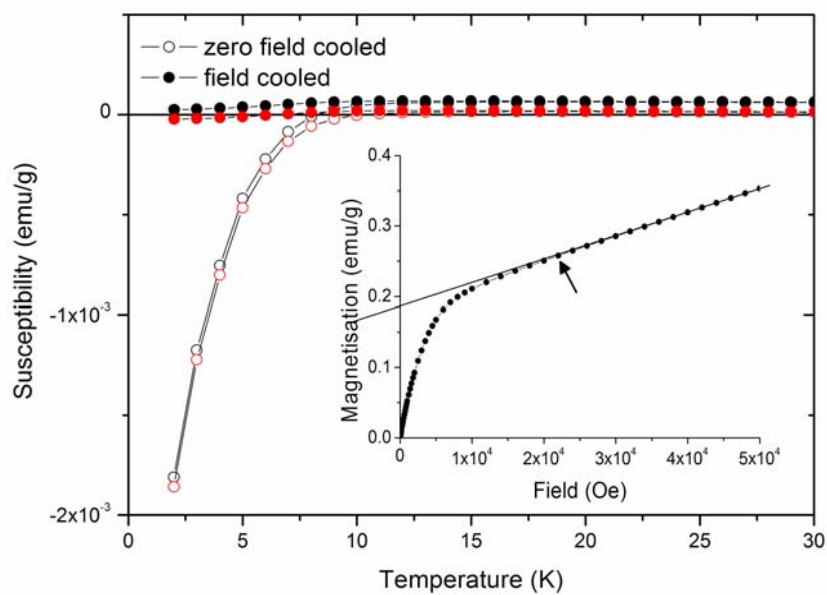

**Fig. S2** Zero-field-cooled (ZFC) (black open circles) and field-cooled (FC) (black closed circles) susceptibilities versus temperature for NaFeAs measured in an applied field of 50 Oe (red open and closed circles show the ZFC and FC curves corrected for a Fe impurity of ~0.08% by mass); Inset: Magnetisation versus applied field of NaFeAs at 300 K.

Correction of the susceptibility data for a small elemental iron impurity. The plot of magnetisation versus applied field at 300 K, (Fig. S2, inset) reveals the presence of 0.08 % by mass of metallic Fe deduced from the saturation field ($H_{sat(Fe)} \approx 2.2 \times 10^4$ Oe) and saturation magnetisation of Fe (222 emu g$^{-1}$). This level of impurity is much smaller than can reasonably be detected by diffraction methods. Correcting for this impurity using the measured susceptibility of iron under the conditions used for measuring the NaFeAs sample (Fig. S2, red symbols) reveals a diamagnetic field-cooled susceptibility at low temperature. Note that this small amount of elemental iron is not sufficient to account for the small value of the diamagnetic signal. The corrected values are used in Figure 3 of the main article.

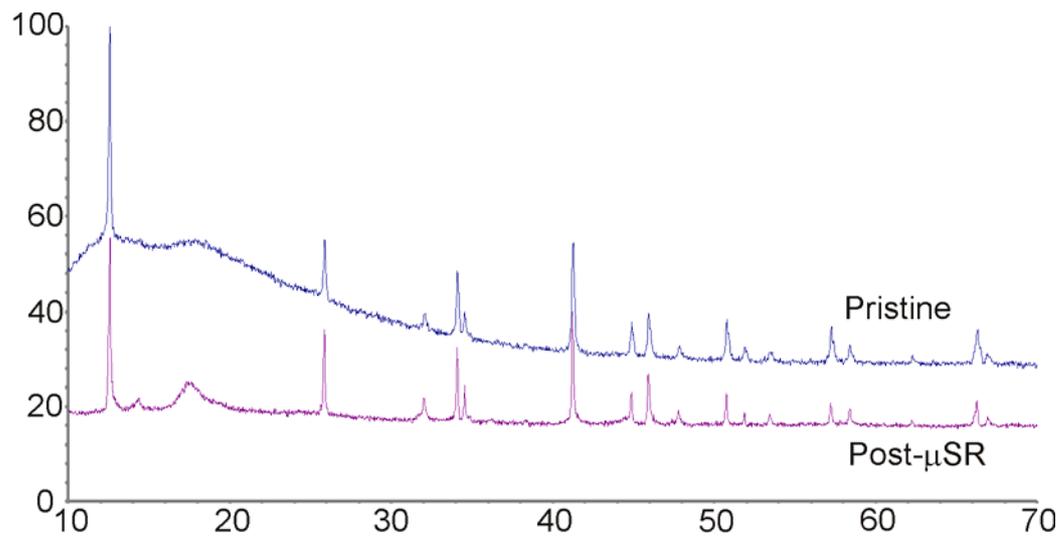

**Fig. S4.** Comparative X-ray diffractograms of the sample in its pristine state and after return to the laboratory after the µSR experiment. The broad features between 15 and 20 degrees in both diffractograms arise from the air-tight sample holder. The susceptibilities are compared in Figure 3 in the main text.